\font\sc=cmcsc10

\def\kpc{\;{\rm kpc} }

\def\Msun{\,M_\odot}

\def\kms{\;{\rm km}~{\rm s}^{-1}  }

\def\vlosi{v_{\odot,i}}
\def\vlos{v_{\odot}}
\def\vlosic{v_{{\rm r}\odot,i}}
\def\vlosc{v_{{\rm r}\odot}}
\def\secthetamom{{\langle v_{\theta}^2\rangle}}
\def\secrmom{\langle v_r^2 \rangle}
\def\rhos{\rho_{\rm s}}
\def\as{a_{\rm s}}
\def\ra{r_{\rm a}}
\def\model{{\rm model}}
\def\data{{\rm data}}
\def\vt{v_{\rm t}}
\def\vc{v_{\rm c}}

\def\spose#1{\hbox to 0pt{#1\hss}}
\def\lta{\mathrel{\spose{\lower 3pt\hbox{$\sim$}}
    \raise 2.0pt\hbox{$<$}}}
\def\gta{\mathrel{\spose{\lower 3pt\hbox{$\sim$}}
    \raise 2.0pt\hbox{$>$}}}

\documentstyle[12pt,aasms4,epsf,rotate]{article}

\begin{document}

\title{Dynamical Mass Estimates for the Halo of M31 from Keck
Spectroscopy\footnote{Based in part on data collected at W.\,M.\ Keck
Observatory, which is operated as a scientific partnership among the
California Institute of Technology, the University of California and the
National Aeronautics and Space Administration.  The Observatory was made
possible by the generous financial support of the W.\,M.\ Keck Foundation.}}
\author{N.\ Wyn Evans}
\affil{Theoretical Physics, 1 Keble Road, Oxford, \\
OX1 3NP, UK \\
Email: {\tt nwe@thphys.ox.ac.uk}}
\author{Mark I.\ Wilkinson}
\affil{Institute of Astronomy, Madingley Road,\\
Cambridge, CB3 OHA, UK \\
Email: {\tt markw@ast.cam.ac.uk}}
\author{Puragra Guhathakurta
\footnote{Alfred P.\ Sloan Research Fellow}}
\affil{UCO/Lick Observatory, Department of Astronomy \& Astrophysics,\\
University of California, Santa Cruz, California 95064, USA\\
Email: {\tt raja@ucolick.org}}
\author{Eva K.\ Grebel
\footnote{Hubble Fellow}}
\affil{University of Washington, Department of Astronomy,\\
Box 35158, Seattle, Washington~98195-1580, USA\\ and \\ Max Planck
Institute for Astronomy, Koenigstuhl 17, \\ D-69117 Heidelberg,
Germany\\ Email: {\tt grebel@astro.washington.edu}}
\and
\author{Steven S.\ Vogt}
\affil{UCO/Lick Observatory, Department of Astronomy \& Astrophysics,\\
University of California, Santa Cruz, California 95064, USA\\
Email: {\tt vogt@ucolick.org}}

\begin{abstract} 
The last few months have seen the measurements of the radial
velocities of all of the dwarf spheroidal companions to the Andromeda
galaxy (M31) using the spectrographs (HIRES and LRIS) on the {\it Keck
Telescope}. This paper summarises the data on the radial velocities
and distances for all the companion galaxies and presents new
dynamical modelling to estimate the mass of extended halo of M31. The
best fit values for the total mass of M31 are $\sim 7 - 10 \times
10^{11} \Msun$, depending on the details of the modelling.  The mass
estimate is accompanied by considerable uncertainty caused by the
smallness of the dataset; for example, the upper bound on the total
mass is $\sim 24 \times 10^{11}\Msun$, while the lower bound is $\sim
3 \times 10^{11} \Msun$.  These values are less than the most recent
estimates of the most likely mass of the Milky Way halo. Bearing in
mind all the uncertainties, a fair conclusion is that the M31 halo is
roughly as massive as that of the Milky Way halo. There is no
dynamical evidence for the widely held belief that M31 is more massive
-- it may even be less massive.

\end{abstract}


\section{INTRODUCTION}

Dynamical studies of the two largest members of the Local Group -- the
Milky Way and the Andromeda (M31) galaxies -- are of particular
interest. The mass of M31 is well constrained on the scales of a few
tens of kiloparsecs from the optical and H\,{\sc i} rotation curves
(Braun 1991) and from the kinematics of the planetary nebulae and
globular cluster populations (e.g., Evans \& Wilkinson 2000). To probe
much larger radii and estimate the total mass of the M31 halo, it is
to the satellite galaxy population that we must turn. Here, there have
been dramatic developments in the last few years. The advent of
wide-field imaging surveys (Armandroff, Davies, \& Jacoby 1998;
Karachentsev \& Karachentseva 1999; Armandroff, Jacoby, \& Davies
1999) has led to the discovery of three new dwarf spheroidals (dSphs)
-- And\,V, And\,VI (Peg\,dSph) and Cas\,dSph. In all, the entourage of
galaxies accompanying M31 now numbers as many as fifteen, although the
status of two of these (Pegasus and IC~1613) is perhaps unclear as
opinions differ as to whether they are close enough to be bound (Mateo
1998; Grebel 1999; Courteau \& van den Bergh 1999).

For the Milky Way, the data set of the eleven companion galaxies, as
well as the distant globular clusters, has often been exploited to
estimate the total mass of the Milky Way halo. The older techniques of
virial estimators have nowadays been superseded by more sophisticated
Bayesian likelihood estimators (Little \& Tremaine 1987). The most
recent analysis by Wilkinson \& Evans (1999) found a total mass for
the Milky Way halo of $\sim 19^{+36}_{-17} \times 10^{11} \Msun$, with
the large error bars reflecting the small sample size and the
uncertainties in the proper motions. Hitherto, the satellites of M31
have received scant attention. Evans \& Wilkinson (2000) looked at all
the available published data on ten of the satellite galaxies and
found a rather low mass of $\sim 12.3^{+18}_{-6} \times 10^{11}\Msun$.
This raises the possibility that the Milky Way Galaxy may be the
largest member of the Local Group.

Here, we exploit recent stellar velocity measurements with the
spectrographs on the {\it Keck Telescopes}. The new observations are
reported in detail by Guhathakurta et al.\ (2000a,b). The results enhance
the dataset on the radial velocities of the companion galaxies of M31
by fifty per cent and are used to provide the most reliable estimate of
the mass of the extended halo of M31 to date.

\section{DATA}

\subsection{Radial Velocity Measurements}

The radial velocities of eight of the companions to M31 have been
known for some time based on one or more of the following techniques:
H\,{\sc i} 21\,cm measurements, optical absorption line spectra of the
integrated light of galaxy cores/nuclei, bulges, or globular clusters,
or optical emission line spectra of H\,{\sc ii} regions.  The data are
recorded in the upper panel of Table~\ref{tab:satdata}.  These
techniques are not easily applicable to galaxies of low surface
brightness and/or low gas content.  Consequently, until very recently,
the radial velocities of the six dSph companions (And\,I, And\,II,
And\,III, And\,V, And\,VI, Cas dSph) remained unmeasured. They require
8--10 meter class telescopes and efficient spectrographs for
spectroscopy of individual red giant stars.  C\^ot\'e et~al.\ (1999)
have recently provided the first kinematic study of And\,II.  They
used the {\it High-Resolution Echelle Spectrometer} (HIRES; Vogt
et~al.\ 1994) on the 10-meter {\it Keck~I Telescope} to measure the
radial velocities of a few red giants in each dwarf galaxy.
Guhathakurta et al.\ (2000a,b) provide the data on the outstanding
five dSphs. Using the {\it Low-Resolution Imaging Spectrometer} (LRIS;
Oke et~al.\ 1995) on the {\it Keck~II Telescope}, they obtained
velocity measurements for nearly a hundred red giants in And\,I,
And\,III, And\,V, and And\,VI. Guhathakurta et~al.\ (2000a,b) have
also carried out HIRES multislit echelle observations of 21~red giant
branch stars in Cas\,dSph.  The radial velocities of the dSphs are
recorded in the lower panel of Table~\ref{tab:satdata}.
          
\subsection{Distance Estimates}

In general, distance estimates for the M31 companions are derived from
variable stars such as Cepheids and RR Lyrae, from the tip of the red
giant branch (TRGB) method (Lee, Freedman, \& Madore 1993), or from
the brightness of horizontal branch (HB) stars (Lee, Demarque, \& Zinn
1994).  One has to rely on TRGB- or HB-based distances for the dSphs,
since they lack the young stellar populations with which Cepheid
variables are associated.

Deep $V$ and $B$ exposures with {\it Hubble Space Telescope}/WFPC2
have been used to clearly detect the HB in And\,I and And\,II
(Da~Costa et~al.\ 1996, 2000).  The distance estimates for these
galaxies are the most accurate of all the M31 dSphs.  The HB method
provides greater distance accuracy than the TRGB method, but HB stars
are faint and difficult to photometer accurately, especially in
crowded fields, and this makes the method impractical for most
ground-based telescopes.  Moreover, the metallicity dependence of the
HB absolute magnitude, $M_V\rm(HB)$, has been the subject of some
debate (cf.  Ajhar et~al.\ 1996).  Distances based on the TRGB method
are available for the rest of the M31 dSph satellites.  Photometry of
bright red giants has been carried out with the KPNO 4-meter
telescope, 3.5-meter WIYN telescope, 6-meter BTA telescope at the
Special Astrophysical Observatory, and with LRIS on {\it Keck~II} for
the following dSphs: And\,III (Armandroff et~al.\ 1993), And\,V
(Armandroff, Davies, \& Jacoby 1998; Grebel \& Guhathakurta 1998),
And\,VI (Grebel \& Guhathakurta 1999; Tikhonov \& Karachentsev 1999;
Armandroff, Davies, \& Jacoby 1999), and Cas\,dSph (Grebel \&
Guhathakurta 1999; Tikhonov \& Karachentsev 1999).  There are several
limitations to the TRGB method: it takes a large sample size to define
the TRGB break accurately; since the method involves the location of a
step in the stellar luminosity function, biases in the distance
estimate are caused by blends and photometric errors; errors in
statistical field subtraction can be a problem; super TRGB populations
such as intermediate-age AGB stars can cause systematic errors.

For all the dSphs, there is substantial error on the distances (see
Table~\ref{tab:satdata}).  Additional uncertainty is introduced
through the poorly known distance of M31 itself, for which a variety
of estimates is available using different methods; we adopt a distance
of 770 kpc.

\section{MASS ESTIMATES}

Let us assume that the halo of M31 is spherically symmetric with the
potential $\psi$ and density $\rho$ given by
\begin{equation}
\rho(r) = {M\over 4\pi}{a^2\over r^2 (r^2 + a^2)^{3/2}},\qquad
\psi(r) = v_0^2\log\Bigl({\sqrt{r^2 + a^2} + a \over r}\Bigr).
\end{equation}
The satellite galaxies are assumed to follow the same density law
$\rhos$ but with the scalelength $\as = 250 \kpc$ (cf., Evans \&
Wilkinson 2000). The velocity normalisation $v_0$ is chosen to
reproduce the circular speed at $30 \kpc$ of $\sim 235 \kms$ (Braun
1991). We adopt two possibilities for the distribution of
velocities. These must depend on the integrals of motion, namely the
binding energy $\varepsilon$ and angular momentum $l$ per unit mass.
The first is a distribution function of form
\begin{equation}
F(\varepsilon, l) = l^{-2\beta} f(\varepsilon),
\end{equation}
so that the orbital anisotropy $\beta = 1 - \secthetamom/ \secrmom$ is
constant. This distribution ranges from radial anisotropy ($\beta >0$)
to tangential anisotropy ($\beta <0$).  The second is a distribution
of Osipkov-Merritt form (see Binney \& Tremaine 1987)
\begin{equation}
F(\varepsilon,l) = f(Q), \qquad Q = \varepsilon - l^2/2\ra^2,
\end{equation}
where $\ra$ is an anisotropy radius. This distribution is isotropic in
the inner parts ($r <\!\!< \ra$) and radially anisotropic in the outer
parts ($r >\!\!> \ra$).  The functions $f(\varepsilon)$ and $f(Q)$ are
written out elsewhere (Wilkinson \& Evans 1999; Evans \& Wilkinson
2000).

Mass estimation using Bayesian likelihood methods was introduced by
Little \& Tremaine (1987). Bayes' theorem states that the probability
of any set of model parameters given the data $P(\model |\,\data)$ is
\begin{equation}
P(\model|\,\data) P(\data) = P(\model) P(\data|\,\model).
\end{equation}
Here, $P(\model)$ describes our prior beliefs as to the likelihood of
the model parameters, while $P(\data|\,\model)$ is the probability of
the data given the model. For the 15 satellite galaxies, the data
consist of three-dimensional positions $r_i$ with respect to the
center of Andromeda, together with their heliocentric line of sight
velocities $\vlosi$. These are corrected for the motion of the Sun
around the Galactic Center and the infall of the Milky Way towards
Andromeda.  We assume a circular speed of $220 \kms$ at the
Galactocentric radius of the sun ($R_{\odot}$ = 8.0 kpc) and a solar
peculiar velocity in $\kms$ of ($U,V,W$) = ($-9,12,7$). The line of
sight velocity of M31, corrected for the motion of the sun, is
$v_{{\rm r,M31}} = -123 \kms$ (Courteau \& van den Bergh 1999).  This
gives the velocities $\vlosic$ in the rest frame of the center of M31.
Along any line of sight, let ($\vt, \eta$) be polar coordinates in the
plane of the sky. Then the required probability is given by
\begin{equation}
\label{eqn:satprob}
P(\data|\model) = \frac{1}{ \rhos (r_i) }
\int_0^{\sqrt{2\psi(r_i)-\vlosic^2}} {{\rm d}}\vt\, \vt
\int_0^{2\pi}{\rm d}\eta\, f(\varepsilon,l)
\end{equation}
The prior probabilities contain the {\it a priori} information about
the model parameters. For the total mass $M$, we use $P(M) \propto {1
/M^2}$, while for the velocity anisotropy, we use $P(\beta) \propto
1/{(3 - 2\beta)^2}$ or $P(\ra) \propto {1/ \ra}$ (see Wilkinson \&
Evans (1999) for a discussion).

The distances to the satellite galaxies are the main uncertainty.  Our
error convolution aims to take account of two factors. First, there is
the quoted error estimate $\Delta s_i$ associated with each distance
given in Table~\ref{tab:satdata}. Second, there is a small, but
distinct, probability $\epsilon$ that some of the published distance
estimates are seriously in error due to systematic uncertainties; we
assume that the probability of a rogue distance $\epsilon$ is
$0.1$. Therefore, we choose our convolution function $E$ to have the
form
\begin{equation}
E(z;i) = (1-\epsilon) B(z;s_i-\Delta s_i,s_i+\Delta
s_i) + \epsilon B(z;s_{\rm min},s_{\rm max}).
\end{equation}
Here, $B$ is the hat-box function
\begin{equation} 
B(d;d_{\rm min},d_{\rm max}) = \left\{ 
\begin{array}{ll}
\displaystyle
\frac{1}{d_{\rm max} - d_{\rm min}} & 
\mbox{$d_{\rm min} < d < d_{\max}$},\\ 0 & \mbox{otherwise},
\end{array} \right.
\end{equation}
while $s_i$ is the published distance estimate of the $i$th satellite
with error $\Delta s_i$ and $s_{\rm min}$ and $s_{\rm max}$ are the
maximum and minimum of the distance estimates listed in
Table~\ref{tab:satdata}.

\begin{figure}
\begin{center} {
               \epsfxsize 0.7\hsize
               \leavevmode\epsffile{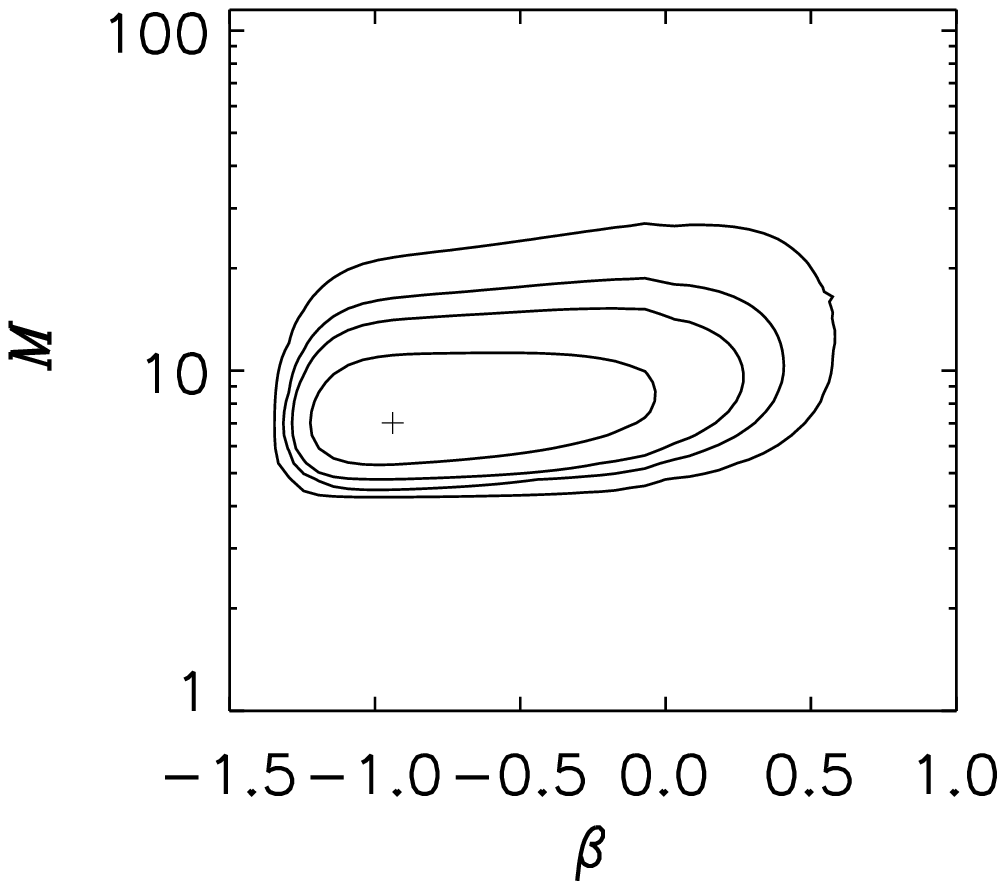}
}
\end{center}
\caption{This figure shows likelihood contours in the plane of total
mass $M$ (in units of $10^{11} \Msun$) and anisotropy parameter
$\beta$. The contours are at heights 0.32, 0.10, 0.045 and 0.01 of the
peak. The most likely mass is $7.0 \times 10^{11} \Msun$ and the most likely
value of $\beta$ is $-$0.90 corresponding to tangential anisotropy.}
\label{fig:figl2b}
\end{figure}

\begin{figure}
\begin{center}
{
              \epsfxsize 0.7\hsize
               \leavevmode\epsffile{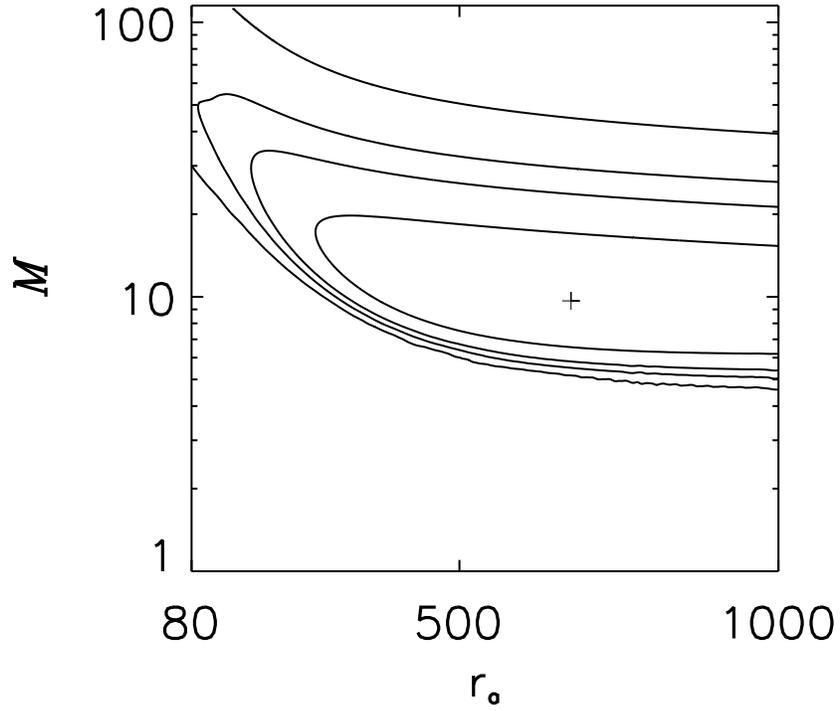}
}
\end{center}
\vspace*{0.5cm}
\caption{This figure shows likelihood contours in the plane of total
mass $M$ (in units of $10^{11} \Msun$) and anisotropy radius
$\ra$. The contours are at heights 0.32, 0.10, 0.045 and 0.01 of the
peak. The most likely mass is $9.4 \times 10^{11} \Msun$ and the most likely
value of $\ra$ is $680 \kpc$.}
\label{fig:figom}
\end{figure}

Figure~\ref{fig:figl2b} shows the results for the constant anisotropy
distributions.  The most likely mass is $7.0 \times 10^{11} \Msun$ and
the most likely value of $\beta$ is $-$0.95 corresponding to
tangential anisotropy.  An earlier analysis found that there is a
tendency to underestimate the mass for small datasets (Evans \&
Wilkinson 2000).  This typically leads to an underestimate of the
order of a factor of two for a dataset of ten to twenty objects, and
is combined with a spread that is of the order of fifty per cent. This
gives a final result of $\sim 7.0_{-3.5}^{+10.5} \times 10^{11}
\Msun$.  The anisotropy is poorly constrained as the likelihood
contours are distended horizontally. However, radial anisotropy is not
favored as the $1 \sigma$ contour is concentrated in the tangentially
anisotropic portion of the figure.  Figure~\ref{fig:figom} shows the
results for the Osipkov-Merritt distribution function. The most likely
mass is $\sim 9.7_{-4.9}^{+14.6} \times 10^{11} \Msun$ and the most
likely value of $\ra$ is $\sim 680 \kpc$, suggesting the velocity
distribution is roughly isotropic.  Table~\ref{tab:othercases} shows
the effects of varying some of our modelling assumptions on the
results for the mass. These include (i) the omission of Pegasus and IC
1613, which may not be bound to M31 (Courteau \& van den Bergh 1999),
(ii) the alteration of the satellite distances to those compiled by
Mateo (1998), (iii) the increase of the error bars on the distances to
$25 \%$, (iv) the alteration of the scalelength of the satellite
galaxy number density distribution and (v) the change of the velocity
normalisation of the halo $v_0$.  The main change is the removal of
Pegasus and IC 1613 from the dataset, which cause a roughly 30\% drop
in the mass estimate.  As can be seen from Table~\ref{tab:othercases},
the other alterations do not have a significant effect on the answer.

\section{CONCLUSIONS}

We have presented the first detailed study of the mass of the halo of
the Andromeda galaxy that uses the radial velocity and distance
measurements for all six dSph satellites (And\,I, And\,II, And\,III,
And\,V, And\,VI and Cas\,dSph).  This takes to fifteen the number of
tracers at large radii and provides a reasonable sample with which to
estimate the total mass. Such calculations have not been possible
before, because of the unavailability of radial velocity data for the
low surface brightness companions of Andromeda. Spectroscopy of
individual red giant stars in these systems has only recently become
feasible thanks to 8--10 meter class telescopes such as {\it Keck} and
state-of-the-art spectrographs (C\^ot\'e et~al.\ 1999; Guhathakurta
et~al.\ 2000a,b).  We reckon that the mass of the Andromeda halo is
$\sim 7.0_{-3.5}^{+10.5} \times 10^{11} \Msun$. This comes from
analysing all the companion galaxies with a family of velocity
distributions of constant orbital anisotropy. The error bars include
the statistical effects of the small dataset, as well as the
uncertainties in the distance estimators.  There is some additional
uncertainty from changes in the modelling assumptions but this is at
the 30\% level at most. For example, altering the velocity
distribution gives us a slightly higher estimate of $\sim
9.7_{-4.9}^{+14.6} \times 10^{11} \Msun$.

The most recent estimate of the mass of the Milky Way is $\sim
19_{-17}^{+36} \times 10^{11} \Msun$ (Wilkinson \& Evans 1999). All
the mass estimates inevitably come with substantial uncertainty, as
they are inferred from the small datasets of satellite galaxies that
are the only known probes of the distant halo. Bearing in mind such
caveats, a fair conclusion is that {\it the Andromeda halo is very
roughly as massive as that of the Milky Way}. There seems little
evidence from the satellite galaxy motions for the widely held belief
that the Andromeda halo is more massive than that of the Milky Way
(e.g., Peebles 1996). Indeed, such evidence as there is points in the
opposite direction, namely that the Andromeda halo may actually be
slightly less massive.

\acknowledgments
EKG acknowledges support by NASA through grant HF-01108.01-98A from
the Space Telescope Science Institute, which is operated by AURA
under NASA contract NAS5-26555.

\eject
\begin{table}[t]
\caption{
Data on M31 and its companion galaxies.  The heliocentric distances s
are taken from Grebel (2000).  The observed line of sight radial
velocities are from Courteau \& van den Bergh (1999; upper panel) or
the new measurements discussed in the text (lower panel).  Listed are
Galactic coordinates ($\ell$,$b$), the heliocentric distances $s$ in
kpc, the observed line of sight radial velocities $\vlos$ in $\kms$,
the distances from the center of M31 $r$ in kpc, corrected line of
sight velocities $\vlosc$ (adjusted for the solar motion within the
Milky Way and the radial motion towards M31), and object type.}
\label{tab:satdata}
\begin{center}
\begin{tabular}{rrrr@{}lr@{}lrrc}
\hline\hline
Name& \multicolumn{1}{c}{$\;\ell$}& \multicolumn{1}{c}{$\;\;b$}&
\multicolumn{2}{c}{$s$}&\multicolumn{2}{c}{$\;\vlos$}&
\multicolumn{1}{c}{$\;r$} 
& \multicolumn{1}{c}{$\vlosc$} & Type\\ \hline
\vspace{-.3cm}\\
M31 & 121.2 & $-21.6$ & 770 &$\pm 40$ & $-301$ &$\pm 1$& $-$& $-$& SbI-II\\
\vspace{-.3cm}\\
M32 & 121.1 & $-22.0$ & 770 &$\pm 40$ & $-205$ &$\pm 3$ & 5
& $+95$ & E2\\
NGC 205 & 120.7 & $-21.1$ & 830 &$\pm 35$ &  $-244$ &$\pm 3$ & 61
& $+58$ & dSph\\ 
NGC 147 & 119.8 & $-14.3$ & 755 &$\pm 35$ &  $-193$ &$\pm 3$ &
100 & $+118$ & dSph/dE5\\ 
NGC 185 & 120.8 & $-14.5$ & 620 &$\pm 25$ & $-202$ &$\pm 7$ &
173 & $+107$ & dSph/dE3\\ 
M33 & 133.6 & $-31.5$ & 850 &$\pm 40$ & $-180$ &$\pm 1$ &
225 & $+72$ & ScII-III\\ 
IC 10 & 119.0 & $-3.3$ & 660 &$\pm 65$ & $-344$ &$\pm 5$ & 253
& $-29$ & dIrr\\ 
LGS3 & 126.8 & $-40.9$ & 810 &$\pm 60$ &  $-286$  &$\pm 4$ &
276 & $-38$ & dIrr/dSph\\ 
Pegasus & 94.8 & $-43.5$ & 760 &$\pm 100$ & $-182$ &$\pm 2$ &
409 & $+86$ & dIrr/dSph\\ 
IC 1613 & 129.7 & $-60.6$ & 715 &$\pm 35$ &  $-232$ &$\pm 5$ &
504 & $-58$ & IrrV\\ 
\vspace{-.3cm}\\
\hline
\vspace{-.3cm}\\
And\,I & 121.7 & $-24.9$ & 790 &$\pm 30$ &  $-380$ &$\pm 2$ &
49 & $-85$ & dSph\\ 
And\,II & 128.9 & $-29.2$ & 680 &$\pm 25$ &  $-188$ &$\pm 3$ & 158 & $+82$ & dSph\\ 
And\,III & 119.3 & $-26.2$ & 760 &$\pm 70$ &  $-355$ &$\pm 10$ &
67 & $-58$ & dSph\\ 
And\,V & 126.2 & $-15.1$ & 810 &$\pm 45$ &  $-403$ &$\pm 4$ &
118 & $-107$ & dSph\\ 
And\,VI & 106.0 & $-36.3$ & 775 &$\pm 35$ &  $-354$ &$\pm 3$ &
265 & $-65$ & dSph\\ 
Cas\,dSph & 128.5 & $-38.8$ & 760 &$\pm 70$ &  $-307$ &$\pm 2$ &
244 & $-57$ & dSph\\ 
\vspace{-.3cm}\\
\hline\hline
\end{tabular}
\end{center}
\end{table}
\begin{table}[b]
\caption{Illustration of the effects of
changing some of the assumed model parameters. The parameter which has
been changed is given in the first column. The second and third
columns refer to distributions of the form $l^{-2\beta}
f(\varepsilon)$ while the fourth and fifth columns are for the $f(Q)$
models. Also recorded are the most likely values of $M$ (in units of
$10^{11} \Msun$), as well as the anisotropy parameters $\beta$ and
$\ra$ (in kpc).}
\label{tab:othercases}
\begin{center}
\begin{tabular}{ccccccccc}
\hline\hline
Comment & Most likely & Most likely &\null& Most
likely & Most likely \\
& $\beta$ & $M$ &\null& $\ra$ & $M$ \\
\hline\vspace{-.2cm}\\
Canonical & $-0.90$ & 7.0 &\null & 680 & 9.4  \\
Pegasus \& IC 1613 & $-0.95$ & 4.8 &\null & 224 & 6.8  \\
omitted & & &\null & &  \\
Sat. distances & $-0.85$ & 8.7 && 659 & 11.2 \\
altered & & &\null& &\\
$\Delta s = 25\%$ & $-0.90$ & 8.1 &\null& 610 & 10.8   \cr
$\as = 150$ kpc & $-0.95$ & 7.6 &\null& 627 & 11.1 \\
$\vc = 270\kms$ & $-0.95$ & 7.0 &\null& 697 & 9.3  \cr 
$\vc = 200\kms$ & $-0.95$ & 7.6 &\null& 648 & 10.2  \\
\vspace{-.2cm}\\
\hline\hline
\end{tabular}
\end{center}
\end{table}

\eject

\end{document}